\documentclass[%
 aip,
rsi,
 amsmath,amssymb,
reprint,%
]{revtex4-1}

\usepackage{graphicx}
\usepackage{dcolumn}
\usepackage{bm}

\usepackage[utf8]{inputenc}
\usepackage[T1]{fontenc}
\usepackage{mathptmx}
\usepackage{etoolbox}

\makeatletter
\def\@email#1#2{%
 \endgroup
 \patchcmd{\titleblock@produce}
  {\frontmatter@RRAPformat}
  {\frontmatter@RRAPformat{\produce@RRAP{*#1\href{mailto:#2}{#2}}}\frontmatter@RRAPformat}
  {}{}
}%
\makeatother
\begin{document}

\preprint{AIP/123-QED}

\title[Tellurium]{Tellurium Spectrometer for ${}^1\text{S}_0-{}^{1}\text{P}_1$ Transitions in Strontium and Other Alkaline-Earth Atoms}
\author{T. G. Akin}
    \email{thomas.g.akin3.civ@us.navy.mil}
\author{Bryan Hemingway}%
%

\author{Steven Peil}
    \email{steven.e.peil.civ@us.navy.mil}
\affiliation{%
Precise Time Department, United States Naval Observatory, Washington, DC  20392
}%

\date{\today}

\begin{abstract}
We measure the spectrum of tellurium-130 in the vicinity of the 461~nm  ${}^1\text{S}_0-{}^{1}\text{P}_1$ cycling transition in neutral strontium, a popular element for atomic clocks, quantum information, and quantum-degenerate gases.  The lack of hyperfine structure in tellurium results in a spectral density of transitions nearly 50 times lower than that available in iodine, making use of tellurium as a laser-frequency reference challenging.  By frequency-offset locking two lasers, we generate the large frequency shifts required to span the difference between a tellurium line and the ${}^1\text{S}_0-{}^1\text{P}_1$ resonance in strontium or other alkaline-earth atom.  The resulting laser architecture is long-term frequency stable, widely tunable, and optimizes available laser power. The versatility of the system is demonstrated by using it to quickly switch between any strontium isotope in a magneto-optical trap and by adapting it to spectroscopy on a thermal beam with a different alkaline-earth atom.
\end{abstract}

\maketitle

\section{\label{sec:intro}Introduction}

There has been an explosion of interest in advancing quantum technology in recent years~\cite{RevModPhys.89.035002}.  Atom- and ion-based systems are promising candidates for future quantum devices, and atomic clocks are one of the best current examples of an isolated, coherent quantum system put to use in a practical application~\cite{RevModPhys.87.637}.   The most significant advances in atom-based clocks and quantum devices are the use of laser cooling and optical transitions, but the laser technology required for these developments is an impediment to creating robust devices that the community seeks.

The laser source itself is not the only obstacle.  Supporting components such as optical isolators, shutters, and modulators introduce certain challenges and inefficiencies. Another important supporting technology is a suitably stable reference for (pre-)stabilization of the laser frequency.  Common techniques that are adequate for laboratory settings are often inadequate to field in more mature systems.  Optical cavities are sensitive to vibrations and do not provide the long-term stability necessary for some stringent applications~\cite{PhysRevA.74.053801}.  Conventional vapor cells are incompatible with alkaline-earth atoms, the basis for narrow optical-transition systems, due to reactions with the cell windows at the temperatures required to produce an adequate vapor pressure~\cite{Taylor:18}.  Concerns with discharge cells include instabilities and limited lifetimes~\cite{doi:10.1063/5.0051228}. 

Iodine (I$_2$) vapor cells, workhorses of frequency stabilization at most of the visible portion of the spectrum due to the many available molecular transitions, are not viable below the dissociation limit of 500~nm.  On the other hand, transitions in molecular tellurium (Te$_2$) extend through the blue and violet bands of the spectrum relevant for broad transitions in alkaline-earth atoms.  Unlike alkaline-earth metals, Te$_2$ does not react with glass windows at high temperatures and works well in traditional vapor cells. 
Frequencies of molecular transitions are fundamentally stable over long times, vapor cells are insensitive to vibrations, and they are not subject to the drawbacks of discharge cells, making Te$_2$ an excellent frequency reference for many applications. In this report we map the uncharted portion of the tellurium spectrum in the vicinity of the strontium $(5s^2)^{1}\text{S}_0-(5s5p)^{1}\text{P}_1$ transition, we discuss the utility of frequency offset locking for making tellurium vapor cells practical as a general frequency reference at blue wavelengths, and we demonstrate the versatility of this tellurium spectrometer using examples with both strontium and calcium.

\section{\label{sec:spectrum}Tellurium Spectrum in Vicinity of Strontium ${}^{1}\text{S}_0-{}^{1}\text{P}_1$ Transition}

Decades ago, spectral lines in the tellurium dimer were instrumental in calibrations for $1\text{S}-2\text{S}$ spectroscopy in positronium~\cite{PhysRevA.34.4504}, muonium~\cite{PhysRevA.36.4115} and hydrogen~\cite{PhysRevA.41.4632}, as well as providing convenient references for stabilizing argon-ion lasers~\cite{Cancio:97}.  In more recent years tellurium has served as a laser frequency reference for qubit or clock systems using Yb+~\cite{Ma:1993ur} and Ba+~\cite{Dutta:16} ions at 467~nm and 455~nm, respectively. For neutral calcium, the tellurium spectra in the vicinity of the $(4s^2){}^1\text{S}_0-(4s4p){}^1\text{P}_1$ laser cooling transition at 423~nm and the $(4s4p){}^3\text{P}_1-(4p^2){}^3\text{P}_0$ excited-state detection transition at 431~nm have been mapped~\cite{Taylor:18}. Yet, despite the growing use of alkaline-earth metals, application of tellurium beyond these examples has been lacking~\cite{repump}.  In particular, strontium is being used for optical clocks~\cite{Nicholson:2015te}, qubits~\cite{Li:2020vp} \cite{PhysRevLett.124.203201} and quantum gases~\cite{PhysRevA.87.013611}, and the spectrum of tellurium relevant to the $^{1}\text{S}_0-^{1}\text{P}_1$ cooling and (ground-state) detection line at 461~nm is of interest.

\begin{figure}
\includegraphics[width=0.5\textwidth]{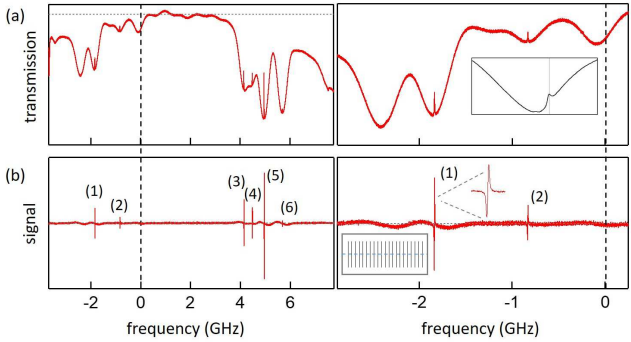}
\caption{\label{f.spectrum}(Color online.)  Tellurium spectrum in the vicinity of the strontium 461~nm $^{1}\text{S}_0-{}^{1}\text{P}_1$ transition. (a) Transmission of probe beam through the vapor cell as a function of frequency detuning from $^{88}$Sr at 650.504~THz.  Doppler-broadened ro-vibrational resonances with some saturation dips visible over a 12~GHz span (left) and more detailed scan of peaks closest to $^{88}$Sr (right).  The strontium resonance with $^{88}$Sr saturation dip (and much smaller $^{86}$Sr dip) shown in inset. (b) Dispersion curves obtained from (a) via FM spectroscopy.  Six Doppler-free signals are observable in this 12~GHz spectral region and labeled as shown (left).  Peaks (1) and (2) shown in more detail, along with a magnified resonance showing a 5~MHz scan of Te(1) (right).  Vertical lines representing 21 iodine hyperfine transitions in the 900~MHz R(37)16-1 ro-vibrational resonance at 519~THz are shown to scale in the inset, bottom left~\cite{Hong:09}.}
\end{figure}

We map out a 12~GHz portion of the tellurium spectrum in the vicinity of the strontium 461~nm line using 10~mW of laser light from an external-cavity diode laser (ECDL), shown in Fig.~\ref{f.spectrum}.  This light is divided between pump and probe beams and applied to a 5~cm-long tellurium-130 vapor cell fabricated by diffusion bonding optical quality quartz windows onto a quartz body. The cell is heated in a bench-top clam shell heater to 500~C. Saturation absorption features can be observed along with the Doppler-broadened ro-vibrational resonances (Fig~\ref{f.spectrum}(a)), and standard frequency-modulation (FM) techniques can be used to generate derivative signals that can be used for laser locking (Fig~\ref{f.spectrum}(b)).  The range of the laser frequency scan is calibrated using a Fizeau wavemeter.

Because of the limited power of ECDLs at 461~nm, independent lasers are used to measure the tellurium and strontium spectra.  The strontium transition is measured by sending 10~mW of laser light transverse to the velocity of a strontium atomic beam in a pump-probe configuration, producing a saturation dip for ${}^{88}$Sr (and a barely perceptible saturation dip for ${}^{86}$Sr at the base of the transmission curve) with the resolution limited by the natural width of the transition (30~MHz).  The 1.25~GHz span of the strontium scan is calibrated with the same wavemeter used for tellurium.  The strontium resonance is shown in the inset of Fig.~\ref{f.spectrum}(a).  

The origin for the $x-$axis in Fig.~\ref{f.spectrum} is the frequency of the ${}^{88}$Sr ${}^1\text{S}_0-{}^1\text{P}_1$ resonance at 650.504~THz.  The ECDL used for strontium spectroscopy is locked to the center of the saturation absorption signal and the ECDL used for tellurium spectroscopy is locked to the closest tellurium peak, labeled (2) in Fig.~\ref{f.spectrum}(b).  We mix these stabilized lasers on a high-speed photodetector and generate a beatnote with a center frequency of 830~MHz, giving the frequency difference of the two resonant frequencies and allowing us to calibrate the origin of the $x-$axis for the Te$_2$ spectrum.

\subsection{\label{sec:spectrometer}Spectrometer}

Iodine-127 is commonly used as a frequency reference from 900~nm to close to 500~nm, where molecular dissociation occurs.  Tellurium-130 picks up where iodine leaves off, providing transitions that can serve as frequency references from 500~nm to below 400~nm.  In addition to lines in the vicinity of the ${}^1\text{S}_0-{}^1\text{P}_1$ strontium transition, the tellurium spectrum is expected to cover this transition in other neutral-atom optical-clock species.

Yet, despite the availability of transitions throughout this wavelength range, the spectral density of lines is low, as can be seen in Fig.~\ref{f.spectrum}.  There, the average separation between useable Te$_2$ lines observed is 2~GHz, with a span of about 5~GHz with no available transitions.  This is in contrast to iodine, where the hyperfine transitions generated by the nuclear spin $I=5/2$ are much more closely spaced, giving a density of useful transitions close to 50 times that available from tellurium.  For example, in the R(37)16-1 (ro-vibrational) transition spanning 900~MHz at 519~THz, there are 21 hyperfine lines available at an average spacing of 40~MHz~\cite{Hong:09}. To appreciate the difference in spectral density of transitions in these two molecules, we add an inset to Fig.~\ref{f.spectrum}(b) illustrating 21 iodine lines equally spaced over 900~MHz. With no nuclear spin, the tellurium spectrum has no density of lines finer than the ro-vibrational splitting, typically on order of 500~MHz to 1~GHz.   Additionally, some of the ro-vibrational lines show no saturation at reasonable power levels as can be seen in Fig.~\ref{f.spectrum}.  As a result, using tellurium as a laser frequency reference for an atomic transition typically involves spanning a frequency difference of hundreds of MHz, up to several GHz.  For the general case, this is not feasible using an AOM due to inefficiencies at such large frequency shifts~\cite{DUTTA2019109}. 
   
\begin{figure}
\includegraphics[width=0.45\textwidth]{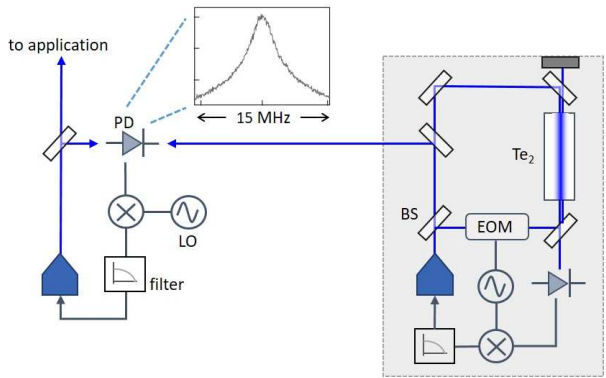}
\caption{(Color online.) Illustration of tellurium spectrometer.  A diode (or other) laser is locked to a tellurium vapor cell using modulation-transfer spectroscopy.  A beatnote generated between this laser and the system laser is mixed to baseband with an RF signal (LO) tuned to the desired offset frequency.   The beatnote shown has a full-width of 2.5~MHz, reflecting the width of the Te$_2$ resonance. The mixed down signal is fed through a loop filter and back to an actuator on the system laser. For applications discussed here, feedback to the PZT suffices. BS - beamsplitter, PD - photodiode, EOM - electro-optic modulator.} \label{f.spect}
\end{figure}

A well-established technique for generating a large frequency shift is to frequency-offset lock one laser to another~\cite{Stace_1998}\cite{doi:10.1063/1.1149573}, as illustrated in Fig.~\ref{f.spect}.  The beatnote generated between two lasers is mixed to baseband with a local oscillator (LO) tuned to the desired offset frequency.  Feeding this mixed-down signal through a loop filter and back to one of the lasers results in a fixed, stable frequency difference.  The technique is versatile in that it can be used for a frequency offset of any value, limited only by the speed of the photodetector, and the offset frequency can be changed quickly, at a rate limited by the servo bandwidth.  Only a very small amount of optical power ($<1$~mW) is required to generate the beatnote with adequate signal-to-noise ratio (SNR) for frequency locking, which is all that is required for these applications.

We lock an ECDL with output power of $>$10~mW to a Doppler-free, 2.5~MHz-wide transition in tellurium as discussed in the previous section.  Less than 1~mW from this laser and from a second laser, of type and power to satisfy requirements of the application at hand, are combined in a fiber splitter to generate a beatnote using a high-speed, fiber-coupled photoreceiver.  Each laser has a linewidth of about one hundred kHz and exhibits drift on order of 10~kHz/s before stabilization to Te$_2$.  An external RF source is used to lock the second laser to the first with a frequency offset that can easily be as large as 10~GHz, making any atomic transition of interest within reach.  The 2.5~MHz width of the locked laser is well below the 30~MHz natural width of the ${}^1\text{S}_0-{}^1\text{P}_1$ strontium transition.

This architecture results in a laser that has nearly all of its output power available for the application of interest, has frequency tuning of several GHz, and has long-term frequency stability provided by the tellurium reference.  The cost of these benefits is a second laser, but this laser can be of modest power ($\sim10$~mW for locking to Te$_2$) and modest spectral quality (linewidth of order 1~MHz for optimal Te$_2$ spectroscopy).

We point out that a technique similar to that presented here has been successfully applied to operational microwave clocks.  Six rubidium fountains in continuous operation for the U.S. Naval Observatory master clock timescale use two ECDLs that are frequency-offset locked with a difference of 6.8~GHz, the ${}^{87}$Rb ground-state hyperfine frequency~\cite{2017}.  The trap laser is locked to a rubidium vapor cell and used for cooling, trapping, launching, state selection and detection.  Because it is impractical to generate the repump light 6.8~GHz away using acousto-optics, a second ECDL is frequency-offset locked to the trap laser.  Nearly all of its power is available for optical pumping during the cooling process and final state read-out.  This approach has been used operationally for more than a decade. 

\subsection{Optical Frequency Reference}

In addition to laser stabilization, a tellurium resonance can be considered for an optical frequency reference.  Atomic and molecular vapor cells have been investigated as candidates for optical clocks for decades, promising improved robustness at the expense of achievable frequency stability compared to cold-atom clocks. Recently, enthusiasm for these technologies has been driven by interest in a new generation of clocks for applications outside of the laboratory. Both two-photon rubidium~\cite{PhysRevApplied.9.014019} and molecular iodine~\cite{PhysRevLett.87.270801} systems have advanced significantly and are being pursued for operations in space~\cite{Schuldt2021}.

The two-photon rubidium resonance has a 330~kHz natural linewidth, and typical linewidths in iodine near 530~nm are 500~kHz or broader. For our operating parameters in tellurium we measure linewidths of about 2~MHz, consistent with other measurements~\cite{Gillaspy:91}. Signal-to-noise ratios of 100 at 1~ms integration time are achievable in these vapor cells. The lower density of states due to the lack of hyperfine structure in tellurium suggests that the signal size should be greater than that for iodine for a given vapor pressure. Additionally, the temperature dependence of the tellurium frequency driven by the pressure shift, reported to be 1~MHz/Torr~\cite{PhysRevA.34.4504}, is similar to that measured in iodine~\cite{Barwood_1984} . As a frequency reference, a tellurium cell is competitive with other vapor-cell systems, particularly if the linewidth can be reduced.

As in most vapor cells, pressure broadening is likely, and considerably narrower tellurium linewidths should be achievable by reducing the cell temperature and compensating with a longer interaction length. Additionally, molecular iodine resonances become more narrow at shorter wavelengths approaching the dissociation limit of $\sim500$~nm, and if the linewidth decreases faster than the line strength, optimal resonances can be found in that part of the spectrum~\cite{Cheng:02}. Such investigations have not been carried out in tellurium, which has a dissociation limit below 390~nm; but with the generation of light at these wavelengths now routine using nonlinear frequency conversion, such an investigation would be possible and of interest. As reported in Ref.[27], hyperfine interactions may complicate the investigation of linewidth versus wavelength, and so tellurium may prove an interesting system for investigating spectroscopy near the dissociation limit.

\section{Applications}

\subsection{Resonance for Different Isotopes}

It is necessary to be able to interact with more than one isotope of a given element in order to exploit the properties of fermions or bosons for a given application. Optical lattice clocks have been demonstrated with both bosonic~\cite{PhysRevA.81.023402} and fermionic~\cite{Nicholson:2015te} isotopes of various optical clock elements, with the specific application dictating which of the tradeoffs between the two options win out.  Some of the most intense interest in quantum degenerate gases involves Bose-Fermi gas mixtures~\cite{PhysRevA.87.013611}, including mixtures with both components in the superfluid state~\cite{doi:10.1126/science.1255380}.

With isotope shifts of the ${}^1\text{S}_0 - {}^1\text{P}_1$ transitions in alkaline-earth and similiar ${}^1\text{S}_0$-ground state atoms of hundreds of MHz, tuning laser frequencies for different isotopes with an AOM is not practical.  Using the tellurium spectrometer introduced above, it is straightforward to produce laser light resonant with any isotope of an element by changing the LO frequency used in the frequency-offset lock. 

\begin{figure}
\includegraphics[width=0.45\textwidth]{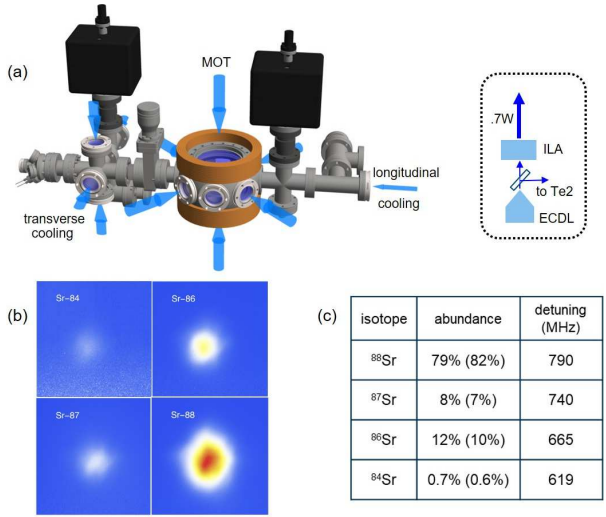}
\caption{(Color online.) (a) Illustration of laser cooling apparatus to trap any naturally occurring isotope of strontium in a magneto-optical trap (MOT).  Examples of different cooling beams are shown, and the laser lock to Te$_2$ is illustrated on the right.  (b) Fluorescence images from four different isotopes of strontium.  Each isotope can be trapped by changing the frequency offset from Te(2) to the corresponding value shown in the table in (c).  The estimated values of the natural abundances are shown in the table along with the established values in parentheses.} \label{f.mot}
\end{figure}

We routinely trap each naturally occurring Sr isotope in a magneto-optical trap (MOT) using the set up shown in Fig.~\ref{f.mot}(a).  The laser cooling apparatus consists of a dispenser to create a strontium beam in a source chamber, separated from the science chamber by a small distance with no intentional differential pumping included.  Modest longitudinal slowing is accomplished using circularly polarized laser light counter-propagating to the atomic beam in the presence of the fringing field from the MOT coils; no explicit Zeeman coil is incorporated.  Two-dimensional transverse cooling is applied in the source region. The MOT coils generate a magnetic-field gradient of 40~G/cm at the science chamber. Six MOT laser beams are created by retro-reflection of three incident beams. Features such as dispenser loading and absence of a Zeeman coil for longitudinal slowing make for a compact system.

All of the cooling laser light is from a 700~mW diode laser system consisting of an ECDL followed by an injection-locked amplifier (ILA).  The 700~mW laser is divided among longitudinal and transverse cooling beams and MOT beams.  The ECDL is frequency-offset locked to a tellurium spectrometer locked to Te(2) in Fig~\ref{f.spectrum}(b).  The frequency difference between Te(2) and the laser frequency at the MOT is set to 790~MHz to trap $^{88}$Sr, determined by optimizing fluorescence from the trapped atoms.  Fluorescence images from $^{84}$Sr, $^{86}$Sr, $^{87}$Sr and $^{88}$Sr atoms are shown in Fig.~\ref{f.mot}(b), and the detunings between Te(2) and the MOT laser required to trap each isotope are shown in the table in Fig.~\ref{f.mot}(c). Integrating over the fluorescence images allows us to estimate the relative abundance of each isotope, listed in the table along with the known values in parentheses.  

The speed at which the laser frequency can be changed is limited only by the bandwidth of the feedback loop, which can easily be greater than 1~MHz if the feedback is applied to the laser current. For our feedback loop, the correction voltage adjusts the laser frequency at a rate of about 100~MHz/ms, driving the largest frequency change required to switch isotopes in several ms. This response time for a change in frequency is much faster than the MOT dynamics, which, in the absence of repumping, are governed by a trapping rate on order of 50~ms.   

The Te$_2$ spectrometer provides a means of trivially switching between Sr isotopes in a blue MOT or similar laser-cooling application.  Isotope shifts of the ${}^1\text{S}_0 - {}^1\text{P}_1$ transition in other alkaline-earth atoms are also easily accommodated with this architecture.

\subsection{Resonance for Disparate Velocities}

Alkaline-earth atomic systems start with an atomic beam from a source heated to 400-600$^{\circ}$C to generate a suitable atomic flux. Some quantum sensors, particularly clocks and interferometers, can exhibit good performance with an uncooled or ``thermal'' atomic beam~\cite{PhysRevLett.78.2046}\cite{PhysRevLett.123.073202}.  In such cases, the $(s^2)^{1}\text{S}_0-(sp)^{1}\text{P}_1$ cycling transition can be used for enhanced detection, producing a signal of many photons per atom (compared to a maximum of 1 photon per atom without a detection laser)~\cite{McFerran:10}.  

Calcium is often used for optical clocks that operate on the singly-forbidden $(s^2){}^1\text{S}_0-(sp)^3\text{P}_1$ transition, as is the case for a thermal-beam clock~\cite{PhysRevLett.123.073202}, due to its narrow natural linewidth of 400~Hz.  The 423~nm $(4s^2){}^1\text{S}_0-(4s4p){}^1\text{P}_1$ transition in calcium lies 380~MHz from the nearest available tellurium line~\cite{Taylor:18}; the laser light for this detection transition can be generated by frequency-offset locking, providing the long-term frequency stability of a molecular transition without sacrificing laser power. 

\begin{figure}
\includegraphics[width=0.45\textwidth]{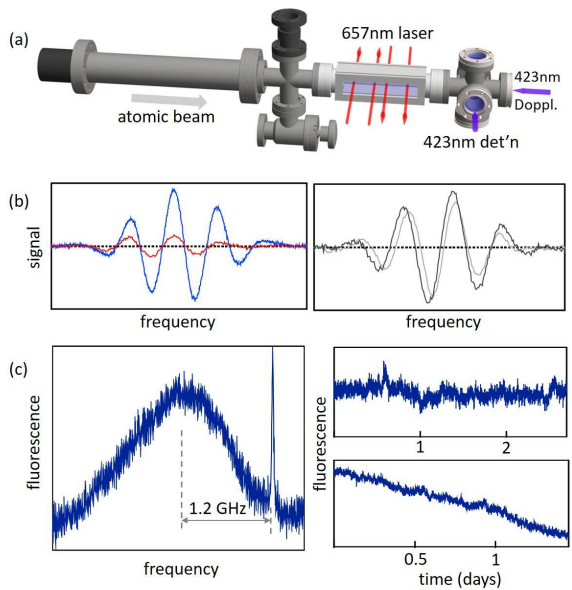}
\caption{(Color online.) (a) Illustration of a thermal calcium beam for an optical clock.  Red (657~nm) spectroscopy laser and blue (423~nm) detection and velocity-monitor lasers are shown. (b) Calcium-clock resonances (interference-fringes obtained with Ramsey-Bord\'{e} spectroscopy~\cite{PhysRevA.30.1836}) for detection using 657~nm photon emission (red) and 423~nm laser induced fluorescence (blue), left. Calcium-clock resonances from two clocks using single detection laser, right. 
(c) Fluorescence from atoms in thermal beam as a function of laser frequency, showing Doppler-broadened (and shifted) distribution from longitudinal laser beam and zero-velocity atoms from detection laser transverse to atomic beam, left. Two examples of behavior of fluorescence from atoms with $v=v_{\text {mp}}$ over time, right.  The top plot shows stable behavior of almost 3 days, while the bottom plot shows a drift which coincides with a drift in vacuum-chamber temperature.}\label{f.calcium}
\end{figure}

A layout for a thermal calcium beam is shown in Fig.~\ref{f.calcium}(a).  An oven generates an atomic beam, which traverses a region where Ramsey-Bord\'{e} spectroscopy~\cite{PhysRevA.30.1836} is carried out with 657~nm laser light.  The outcome of the interaction can be determined by emission of a single 657~nm photon per atom, or by the scattering of many photons per atom by driving the 423~nm detection transition~\cite{McFerran:10}.  Resonances obtained with each technique and the improvement in signal-to-noise ratio (SNR) are shown in Fig.~\ref{f.calcium}(b). With negligible loss of optical power required to generate the frequency lock, the detection laser can divided among multiple clocks simultaneously, reducing the number of laser systems needed (Fig.~\ref{f.calcium}(b)).  This is particularly relevant for timing institutions that operate many atomic clocks but want to minimize the total number of lasers used.

Precise clocks and other sensors using a thermal beam are sensitive to changes in second- and residual first-order Doppler shifts brought about by changes in the velocity distribution of the atoms that can arise from environmental fluctuations.  A typical most-probable velocity $v_{\text {mp}}$ of 500~m/s has a corresponding Doppler shift of 1.2~GHz and spectral width of several hundred MHz.  The velocity distribution can be well characterized by atomic fluorescence from the $(s^2)^{1}\text{S}_0-(sp)^{1}\text{P}_1$ transition induced by a counter-propagating laser scanned over the velocity distribution.  Fig.~\ref{f.calcium}(c) shows fluorescence as a function of laser frequency when both a counter-propagating beam and a transverse (detection) laser beam are present.  With the tellurium spectrometer, the same laser used for detection can be used to monitor changes in the beam velocity by measuring fluorescence from atoms with $v=v_{\text {mp}}$.  In Fig.~\ref{f.calcium}(c), two examples of the behavior in this fluorescence over time are shown.  The top example shows stable behavior over several days, while the bottom plot shows a drift that turned out to be correlated with a drift in temperature of the vacuum chamber.  Additional velocity information is obtained by dithering the laser frequency and extracting the peak of the Doppler distribution.  In addition to being used in real time to improve clock performance, this technique was also used separately to study the stability of the second-order Doppler shift of the atomic beam.

The versatility of the Te$_2$ spectrometer enables the large detunings required to characterize properties of a thermal atomic beam to be reached with the same laser used for detection of zero-velocity atoms.  The efficient use of optical power opens the possibility of using a single laser for multiple optical beams.

\section{Conclusions}

We have presented a versatile laser architecture based on tellurium spectroscopy and frequency offset locking that is long-term frequency stable, can be effortlessly tuned over many GHz, and preserves the output power of the laser.  The added complexity of introducing a second laser to the system, typically an ECDL or similar that is devoted to tellurium spectroscopy, is comparable to other solutions that would be required, such as an optical amplifier to compensate for loss in power if using an AOM.  Applications include any system requiring significant optical power to drive a $(s^2)^{1}\text{S}_0-(sp)^{1}\text{P}_1$ transition in an alkaline-earth atom, particularly those requiring large frequency changes.  Achieving resonance for any isotope and for any velocity in a broad distribution are relevant examples for state-of-the art quantum technology.

\begin{acknowledgments}
This work is supported by Department of Navy Research, Development, Test and Evaluation funds provided for USNO's Clock Development program.
\end{acknowledgments}

\bibliography{tellurium}

\end{document}